\documentclass[preprint,12pt]{elsarticle}

\newcommand{\fS}{\ensuremath{\mathbf{S}}}

\newcommand{\fM}{\ensuremath{\mathbf{M}}}

\newcommand{\fez}{\ensuremath{\mathbf e_{z}}}

\newcommand{\mean}[1]{\ensuremath{\left\langle#1\right\rangle}}

\usepackage{amssymb}

\journal{Journal of Magnetism and Magnetic Materials}

\begin{document}

\begin{frontmatter}



\title{Implications of a temperature-dependent magnetic anisotropy for superparamagnetic switching}


\author[a]{Martin Stier}
\author[b]{Alexander Neumann}
\author[b,c]{Andr\'e Philippi-Kobs}
\author[b]{Hans Peter Oepen}
\author[a]{Michael Thorwart}

\address[a]{I. Institut f\"ur Theoretische Physik, Universit\"at Hamburg, 
Jungiusstra{\ss}e 9, 20355 Hamburg, Germany}
\address[b]{Institut f\"ur Nanostruktur- und Festk\"orperphysik, Universit\"at
Hamburg, Jungiusstra{\ss}e 11, 20355 Hamburg, Germany}
\address[c]{Deutsches Elektronen-Synchrotron DESY, Notkestr. 85, 22607 Hamburg, 
Germany}

\begin{abstract}
The macroscopic magnetic moment of a superparamagnetic system has to overcome an
energy barrier in order to switch its direction. This barrier is formed by  
magnetic anisotropies in the material and may be surmounted 
typically after $10^9-10^{12}$ attempts per second by thermal
fluctuations. In a first step, the associated switching rate may be described
by a N{\'e}el-Brown-Arrhenius law, in which the energy barrier is assumed as
constant for a given temperature. Yet, magnetic anisotropies in general
depend on temperature themselves which is known to modify the 
N{\'e}el-Brown-Arrhenius law. We illustrate quantitatively the implications 
of a temperature-dependent anisotropy on the switching rate and 
in particular for the interpretation of the prefactor as an attempt frequency. 
  In particular, we show that realistic numbers for the attempt 
  frequency are obtained when 
 the temperature dependence of the anisotropy is taken into account.
\end{abstract}


\end{frontmatter}

\section{Introduction}

The development of devices for magnetic storage media faces several challenges in the 
ongoing miniaturization of information units. One of the fundamental physical problems is 
the so-called superparamagnetic limit. In the small scale limit, the alignment
of the macroscopic magnetic moment of the 
ferromagnetic particle with a direction along a preferred axis is no longer stable, but
permanently affected by thermal fluctuations. Despite similarities between para- and
superparamagnetism, distinct differences exist. Quantum
mechanics allows us to describe paramagnetism rigorously with the magnetic field
treated as a small perturbation. In superparamagnetism, the spins of a
nanostructure are coupled by the exchange interaction which causes a
quasiclassical behavior of a collective magnetic moment with a large variety
of possible energetically continuous states. 
Due to the large number of coupled spins which form the macroscopic magnetic
moment, smaller contributions of the single spins to the total energy may become
important. They can be rooted in, e.g., dipolar and/or spin-orbit interaction
effects. The collection of such secondary effects is commonly summarized to form 
the anisotropies. 
The energy scale of a nanomagnet with its small dimensions can become 
comparable with the thermal energy. The total energy
of the nanomagnet with its collectively formed magnetic moment reveals
remarkable differences compared to the case of a single moment. In particular,
due to the competition with thermal energies, important collective features 
strongly change as function of temperature. For example, 
the blocking of the macrospin along a certain direction (easy axis
of magnetization) is substantially influenced. Here, two energetically
degenerate states are oriented parallel to the easy axis and are separated by 
an energy barrier $\Delta E$. The latter suppresses thermal
switching. In the case of a uniaxial system, the barrier height scales with the
volume $V$ and strength of the anisotropy $K$. At high enough temperature, the barrier
can be overcome and the collective magnetic moment can flip back and forth, 
resulting in a vanishing time averaged magnetic moment. In superparamagnetic
systems, a problem becomes immediately obvious since temperature determines both the
collective magnetization as well as the switching rate. This is in the focus of
the present work. 

Superparamagnetic behavior of nanomagnets is usually analyzed in terms of the
switching of the magnetization
\cite{rohart2010,neumann2014}. To
this end, a switching frequency $f(T)$ is
determined as a function of temperature $T$ which commonly reveals an
Arrhenius-like temperature dependence following  
\begin{equation}
  f(T) = f_0 e^{-\frac{KV}{k_B T}}\, ,\label{eq::arr}
\end{equation}
with $f_0$ being the so-called attempt frequency and $k_B$ the Boltzmann
constant. To obtain an Arrhenius law, a temperature-independent anisotropy
energy has to be assumed, such that a separation of time scales is possible\cite{Kramers1940}.
Thus, the switching events have to be rare compared to the frequent attempts made until switching occurs,
meaning $f(T) \ll f_0$  or $KV\gg k_B T$, respectively. 
When the logarithm of the temperature-dependent
switching frequency is plotted versus the inverse temperature, a
straight line results, with the slope yielding the anisotropy energy and the
intercept yielding the attempt frequency. Such an analysis is common to many
experiments and is appealing due to its simplicity, albeit detailed
investigations whether a clear separation of time scales is really given are
not properly made and are sometimes not even possible due to experimental
constraints. 

Even though the magnitude and the dependence on temperature of the attempt
frequency have been modeled on the basis of detailed assumptions about the
reversal mechanisms 
\cite{brown1963thermal,neel1949,Kramers1940,coffey2012thermal,haenggi1990},
large deviations from these predictions by several orders of magnitude  are
reported in the
literature \cite{krause2009magnetization,dormann1999,dormann2007magnetic}.
Several possible physical explanations for these substantial discrepancies have
been offered since then 
\cite{rohart2010,braun1994nonuniform,braun1993thermally}.

In this work, we readdress these discrepancies by analyzing
results from such an Arrhenius fit. Certainly, the Arrhenius plot itself
is a powerful tool and its validity is confirmed by innumerable successful
applications. The mere numbers that come out of such an analysis are correct on
their own, but point out that the assignment to specific material
properties has to be done with care. In particular, precise knowledge about the
temperature dependence of the exponent in Eq.\ (\ref{eq::arr}) is necessary for
reliable extractions of material parameters. Likewise, a straightforward 
extrapolation of the Arrhenius plot to zero temperature in order 
to derive the prefactor $f_0$ and interpret it as a constant ``attempt frequency'' has to be done 
with caution. 

In this context, it was shown that the use of the free energy instead of the total
energy  entering for the energy barrier 
generates a temperature dependent contribution in connection with  
the entropy \cite{nowak2005spin,skomski1993giant}. This additional contribution 
to the activation energy can reduce the prefactor \cite{skomski} compared 
to Brown's result \cite{brown1963thermal} and plays an important role for large statistical 
ensembles with a broad distribution of activation energies and high-dimensional 
energy landscapes \cite{skomski}. 

Of more importance is the explicit dependence of the anisotropy on temperature,
as addressed in the present work in greater detail. This has been investigated
mostly in the blocked regime where switching is largely suppressed
\cite{fernandez2005,yoon2012temperature,usov2011numerical,he2005}, but has also
been mentioned to influence the superparamagnetic switching  
\cite{fernandez2005,koetzler2003}. 

There are several origins of a temperature dependence of the total anisotropy.
At first, the magnetic anisotropy can vary with temperature due to slight
changes of the structure and stress in the material. This is well known for bulk
materials like Co. Second, the effect can be due
to the shape anisotropy which is often the origin of the uniaxial anisotropy of 
nanoparticles. As this part is determined by the saturation magnetization, it
changes with varying temperature. The third influence originates from the
scaling property of the magneto-crystalline anisotropy with the saturation magnetization in a
power-law like characteristics
\cite{pap1,pap2,callen1965probability,callen1966,wolf1957,skomski2}. 
 The temperature dependence of the effective 
anisotropies in magnetic nanoparticles with different shapes with cubic or uniaxial bulk 
anisotropy and N{\'e}el surface anisotropy has been calculated by using a constrained Monte 
Carlo approach \cite{yanes}. The impact of thermal magnon excitations on coercivity 
has been investigated in Ref.\ \cite{ieee}.

The effect of the temperature dependence of the shape anisotropy on the 
coercivity for aligned Stoner-Wohlfarth systems has been considered 
in Ref.\ \cite{he2005}. The standard N{\'e}el-Brown formula for the coercive field 
has been extended to include the temperature dependence of the magnetization, leading 
to an effective temperature dependent anisoptropy barrier.  The role of a temperature dependent 
magnetocrystalline anisotropy on the coercivity of nanostructured materials was 
investigated theoretically in Ref.\ \cite{fernandez2005}. Furthermore, it was shown in Ref.\ 
\cite{polewko} that the temperature dependence of the magnetic anisotropy also needs to be 
carefully taken into account when the magnetic remanence is considered. 

Experimental results on Co-Fe magnetic nanoparticles \cite{torres} show that the 
temperature dependence of the anisotropy has to be taken into account for a matching 
with a N{\'e}el-Brown-Arrhenius law. In this case, an empirical Br\"ukhatov-Kirensky ansatz 
was used for the temperature dependence of the anisotropy. Such an approach reproduces 
realistic zero-temperature values $K(0)$ for the bulk anisotropies as well as realistic 
times for the inverse attempt frequencies. 

The purpose of this work is not to address the details of the physical effects, which
contribute to the temperature dependence of the anisotropy. Instead, our focus 
is on its general impact on the analysis of the superparamagnetic behavior via
its switching characteristics. We demonstrate that the temperature
interval (which is usually somewhat limited in experiments) in which a finite
number of data points are fitted to an Arrhenius law has to be selected with
care. It may determine very sensitively the resulting parameters  
 of the Arrhenius plot, being the slope (anisotropy) and the intercept (attempt 
frequency). In general, besides the blocking
temperature $T_B$ in superparamagnetism, a second temperature scale becomes relevant in 
the macrospin
description, i.e., the Curie temperature $T_C$, being decisive for the 
magnetic ordering. Thus, experimentally determined prefactors 
\cite{krause2009magnetization,tadic2014magnetic,neumann_phd,tackett2010evidence} have to be carefully interpreted, particularly, but not
exclusively, when the temperature range, in which the switching measurement are
performed, is comparable to the magnetic ordering temperature  \cite{fernandez2005}. 
We illustrate quantitatively 
that naively assuming the applicability of the Arrhenius law may give rise to unintended
misinterpretations if a temperature-independent height of the energy barrier is presumed. 

\begin{figure}
\centering
\includegraphics[width=.65\textwidth]{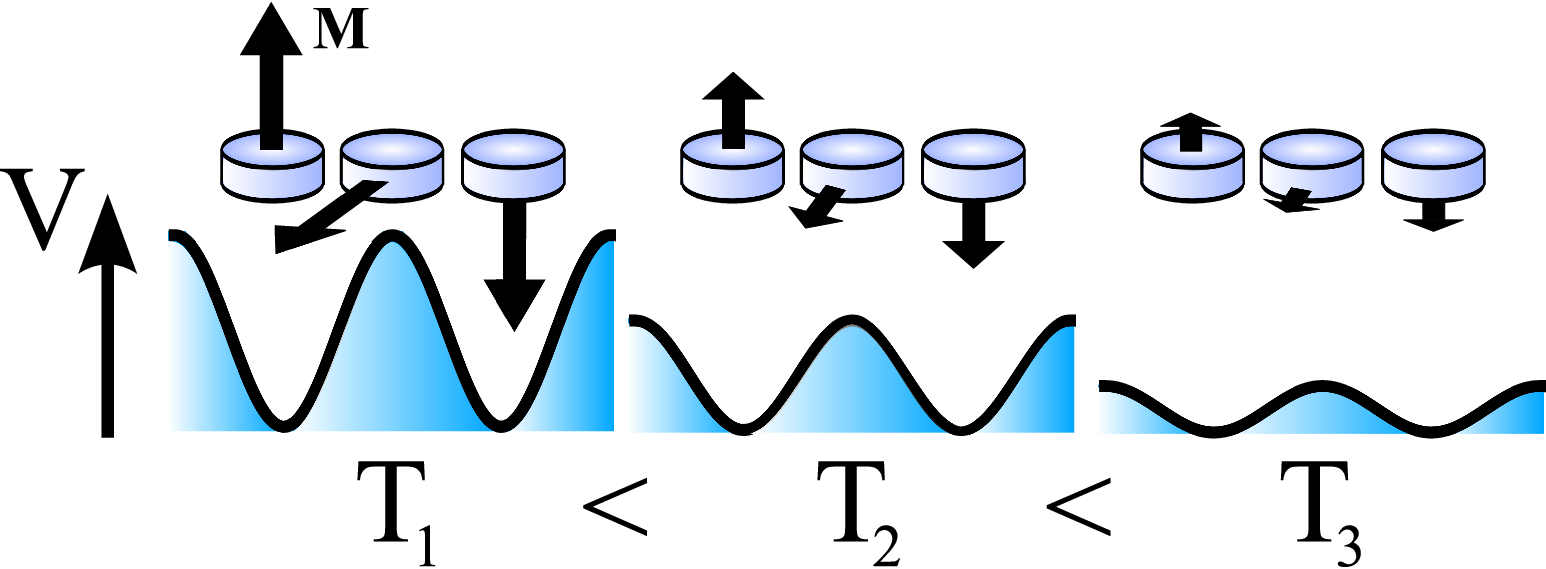}
\caption{\label{fig::sketch}Sketch of magnetic
potentials of superparamagnetic nanomagnets for
different temperatures with the easy axis pointing in the out-of-plane
direction. To switch between the minima, the magnetization
$\mathbf M$ has to overcome the anisotropy energy barrier which typically
depends on $\mathbf M(T)$. With increasing temperature, both the magnetization
and the energy barrier, are decreased.}
\end{figure}

\section{Theoretical models for a temperature-dependent anisotropy}

A generic superparamagnetic system can be described by the Heisenberg model
\begin{equation}
 H= -J\sum_{\mean{i,j}} \fS_i\cdot\fS_j + H_{\rm aniso},\label{eq::ham}
\end{equation}
where the spins $\fS_i$ tend to align parallel due to the exchange interaction
of strength $J$ in order to form a macroscopic magnetic moment. The
exchange interaction itself is rotationally invariant and thus does not favor 
a certain direction for the resulting magnetic moment. It basically determines
the 
temperature dependence of the bulk magnetic system. However, 
 superparamagnetic systems typically possess either intrinsic material
symmetries or external ones such as a shape anisotropy. Then, distinct
directions of the magnetic moment are preferred. To model this feature, a generic
anisotropy term $H_{\rm aniso}$ is introduced in Eq.\ (\ref{eq::ham}). The
specific 
form of the anisotropy has to be defined for the particular experimental
situation. It is important to realize that the anisotropy part in small systems
does not only define easy 
or hard axes, but also influences the temperature dependence of the
magnetization $M(T)$ 
itself. Thus, $M(T)$ in small systems can be very different from comparable bulk 
systems. The theoretical determination of $M(T)$ in those small samples is a non-trivial 
task \cite{nowak2005spin} and depends on the details of the Heisenberg model. As a matter of
fact, we focus on general 
consequences of a temperature dependence of the magnetization and we do not aim
to calculate 
$M(T)$ from first principles. For the purpose of this work, it is sufficient to 
 assign some generic behavior to it. First, we assume the 
magnetic system to be large enough to form a collective macroscopic
magnetic moment but
still sufficiently small to avoid a separation into multiple magnetic domains.
In this superparamagnetic limit, we can replace the collection of microscopic
quantum spins by a classical magnetic moment $\mathbf M$.
Changing to the magnetization as the order parameter, the uniaxial
anisotropy is described by an energy density. The anisotropy part in Eq.\
(\ref{eq::ham}) becomes 
\begin{equation}
 H_{\rm aniso}= -KV\cos^{2}\theta , \label{eq::hamclass}
\end{equation}
with $\theta$ being the angle which the magnetization encloses with the z-axis.
The total Hamiltonian for a
particle of volume V and with the magnetization $\fM$ has two energy minima for
$\fM=\pm |\fM| \fez$.
In order to reverse its direction, the total magnetic
moment has to switch from one minimum to the other by surmounting the
energy barrier $\Delta E = H_{\rm aniso}(\theta=90^{\circ}) -
H_{\rm aniso}(\theta=0^{\circ},180^{\circ}) = K V$ while crossing the highest energy state
at $\theta=90^{\circ}$. The interaction part of Eq.
(\ref{eq::ham}) is irrelevant for determining the energy barrier as it is 
rotationally invariant and will only be considered implicitly as the mechanism
causing the magnetic order. At finite
temperature, the difference of the free energy $\Delta F(T) = \Delta
E(T)-T\Delta S(T)$ determines the height of the switching barrier. However, we
assume that the entropy $S(T)$ does not change significantly by a rotation of
the magnetization and set $T\Delta S\ll \Delta E$\cite{mryasov2005temperature}.

For the following discussion, it is crucial to note that 
the anisotropy is related to the particle magnetization via a conventional power law $K= 
\tilde K[M(T)/M(0)]^{\nu }$ with the temperature dependent  magnetization $M(T)$
and $M(0)$ and \(\tilde K\) being the corresponding values at zero Kelvin 
\cite{nowak2005spin,callen1966,mryasov2005temperature,boetcher2011,
farle1998ferromagnetic}. For the sake of concreteness, we
consider the case where the uniaxial anisotropy is of dipolar origin. Then, 
the energy
difference between hard and easy axis is given by $\Delta E=  \Delta N V M
^{{2}}$ \cite{dubowik1996} with $\Delta N$ stemming from the demagnetization
tensor $\hat N$. Thus, we find 
\begin{equation}\label{enbar}
 \Delta E (T) =\Delta N V M(T)^2\equiv K_{0}M(T)^2\, , \label{eq::dE}
\end{equation}
with $\Delta E(T)$ typically decreasing with increasing temperature. This 
is schematically shown in Fig.\ \ref{fig::sketch}.  
\begin{figure}[t!]
 \centering
 \includegraphics[width=.55\textwidth]{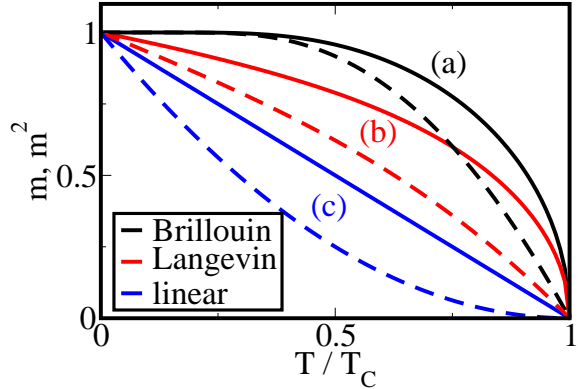}
 \caption{\label{fig::mag}Temperature dependence of
the magnetization $m = M(T) / M(0)$ (full lines) and its square $m^2$
(dashed lines) described by (a) the Brillouin function $B(S,T_C)$ for a 
spin $S=1/2$, (b) the Langevin function $L(T_C)$ for a classical spin ($S\to\infty$), and 
(c) a
linear dependence $m(T) = 1-T/T_C$. 
$T_C$ denotes the Curie temperature of the system. }
\end{figure} 
%

\section{Results and discussion}
\subsection{Generalized N{\'e}el-Brown-Arrhenius law}

Using this model of a temperature dependent energy barrier, we next study the
implication on the switching rate of Eq.\ (\ref{eq::arr}). Therefore, we insert 
Eq.\ (\ref{enbar}) into Eq.\ (\ref{eq::arr}) and find a 
generalized N{\'e}el-Brown-Arrhenius law
\begin{equation}
 f(T) = f_0 e^{-\frac{K_0 M(T)^2}{k_B T}} \, .\label{eq::f}
\end{equation}
In the following, we consider three different conventional model systems
yielding a particular $M(T)$ which actually follow this law and, most
importantly, investigate the consequences of treating the switching dynamics on
the basis of a ``conventional'' Arrhenius law with a temperature-independent
energy barrier, as it is often done in the analysis of experimental data. 
To fix the time scale, we set the attempt frequency at a typical value
of $f_0=10^9$ s$^{-1}$ and choose the anisotropy 
energy as $K_0 M(T=0)^2=1$ eV throughout this work. The conclusions
 below hold even for reasonable deviations from Eq.\ (\ref{eq::f}). For example,
a different exponent $\nu$ in
$M(T)^{\nu}$ may be absorbed in the actual temperature dependence of $M(T)$, while a
(non-exponential) temperature dependence of the prefactor $f_0$
\cite{coffey2012thermal} is in first order negligible.

The actual magnetization of a physical system is affected by a multitude of
material properties, such as intrinsic interactions or the finite sample size. 
While in ferromagnetic bulk samples the temperature dependence of the
magnetization is typically well described by a Brillouin function, this is often
not true in systems of reduced dimensions 
\cite{farle1998ferromagnetic,pierce2004ferromagnetic,
gradmann1974ferromagnetism}. Thus, an assumption on the
general form of $M(T)$ cannot be made without explicitly knowing the specific
system. Yet, to illustrate that the explicit temperature-dependence 
of $M(T)$ has a major effect on the switching rate, we consider three
different examples in this work: (a) the Brillouin function describes 
quantum spins in a bulk sample, (b) the Langevin function refers to a classical
magnetic moment, and (c) a linear dependence such that 
$M(T)=M(0)(1-T/T_C)$ (with the Curie temperature $T_C$) as a possible
variant of an unconventional magnetization curve which can be realized in
magnetic nano-islands \cite{nowak2005spin}. These functions and their squares
are shown in Fig.\ \ref{fig::mag}.

\begin{figure}[t!]
 \centering
 \includegraphics[width=.55\textwidth]{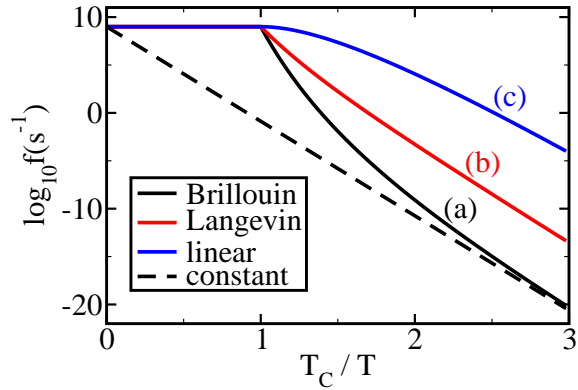}
 \caption{\label{fig::f_for_diff_mag}Decadic logarithm of the switching
rate Eq. (\ref{eq::f}) for the
temperature-dependent magnetizations shown in Fig.\ \ref{fig::mag}. Strong
deviations from an Arrhenius law with a constant anisotropy
$K_0M(T)^2\equiv K_0M(0)^2$ (dashed line) appear even for $T\ll T_C$. Parameters are
$f_0=10^9$ s$^{-1}$ and $K_0 M(0)^2=1$ eV.}
\end{figure}
In general, $M(T)$ decreases with increasing temperature and vanishes at the
Curie temperature $T_C$ which also holds for $\Delta E(T)\propto M(T)^2$.
Thus, the system does not only switch faster with growing temperature due to an
explicit increase of thermal fluctuations, but also due to a lower
anisotropy barrier. This has a pronounced effect on the switching rate which is
shown in
Fig.\ \ref{fig::f_for_diff_mag}. 
For convenience we use the decadic logarithm $\lg\equiv\log_{10}$ to directly
determine the exponent $y$ of the frequency according to
$f=10^y$ s$^{-1}$. Strong deviations from an Arrhenius law with a
constant energy barrier are clearly visible. 

These deviations are not captured by a linear adjustment of
the Arrhenius law with a temperature shift $T\to T-T_0$, 
which yields to the Vogel-Fulcher law. Above $T_C$, this is obvious since the
magnetization, and with it the energy
barrier, vanishes. Close to $T_C$, the explicit values of $f(T\approx T_C)$
lose their physical meaning since the concept of a switching rate with a
separation of time scales breaks down 
\cite{haenggi1990}.  However, from the
existence of a finite critical temperature $T_C$ with $\Delta E(T\ge T_C)\equiv
0$ and $f(T\ge T_C)\equiv f_0$, we have to conclude that such a system cannot 
 be described by a constant energy barrier as in this case $f_{\Delta
E=\rm{const}}(T)$ would approach $f_0$ only for $T\to\infty$. Thus, an
accurate description of the switching rate requires in principle  
the knowledge of the full temperature-dependent magnetization $M(T)$ or,
more precisely, of the energy barrier $\Delta E(T)$. 

\subsection{Interpretation of the prefactor}
The generalized N\'eel-Brown theory of the switching rate has profound 
consequences for the interpretation of experimental data. In
particular, the interpretation of the prefactor in terms of an attempt
frequency has to be revisited. In real systems, it is not a realistic
task to measure the switching rate $f(T)$ over the entire temperature regime from  
$T=0$ to $T_C$. Usually, only a finite (and small) temperature interval
$[T_{\rm{min}},T_{\rm{max}}]$, with $\Delta T =
T_{\rm{max}}-T_{\rm{min}}\approx 30\dots50$ K
\cite{krause2009magnetization,neumann_phd} is experimentally accessible.
 Outside of this range, the switching is either too
fast to be observable with present day tools or too slow to be measured 
in a reasonable time. In fact, $\Delta T$ is usually small enough that the
magnetization does not significantly change with temperature in this selected
temperature window and the switching rate appears to be linearly dependent on
$T^{-1}$ in this small interval. Thus, it is tempting to
fit the results to a linear function 
\begin{equation}
 \lg f = \lg f_0^{\rm{fit}}-\lg(e)\frac{K_0^{\rm{fit}}}{k_B
T}\label{eq::ffit}
\end{equation}
and to determine $f_0$ and $K_0$. However, the
interpretation of the results for the fitted parameters $f_0^{\rm{fit}}$ and
$K_0^{\rm{fit}}$ can actually be misleading. We illustrate this for a
simple analytical example for which we assume at this point a
temperature-dependent magnetization $M(T)=M(0)\sqrt{1-T/T_C}$. The switching
rate resulting from Eq.\ (\ref{eq::f}) is now given by 
$\lg f =\lg f_0^{\rm{fit}} - \lg (e)(K_0 M(0)^2)/(k_B T)$ with  
$\lg f_0^{\rm{fit}} = \lg f_0 + \lg(e)(K_0 M(0)^2)/(k_B T_C)$.
This equation is in fact linearly depending
on $1/T$ and would actually yield the result for $K_0^{\rm{fit}}=K_0M(0)^2$.
However, more importantly, 
the exponent of the fitted prefactor (which is often interpreted as the attempt
frequency) differs from the physical one by $\lg(e)K_0 M(0)^2 /(k_B T_C)$.
This difference can be very pronounced. For exapmle if
$f_0=10^9$ s$^{-1}$ and $K_0 M^2(0) = 1$ eV then
$ \lg f_0^{\rm{fit}}\approx \lg f_0 + \frac{5000\, \rm{ K}}{T_C}$. Hence, 
the attempt frequency $f_0=10^y$ is in this case is
overestimated by five orders of magnitude for a typical $T_C$ of 1000 K.

\begin{figure}[t!]
\centering
 \includegraphics[width=.55\textwidth]{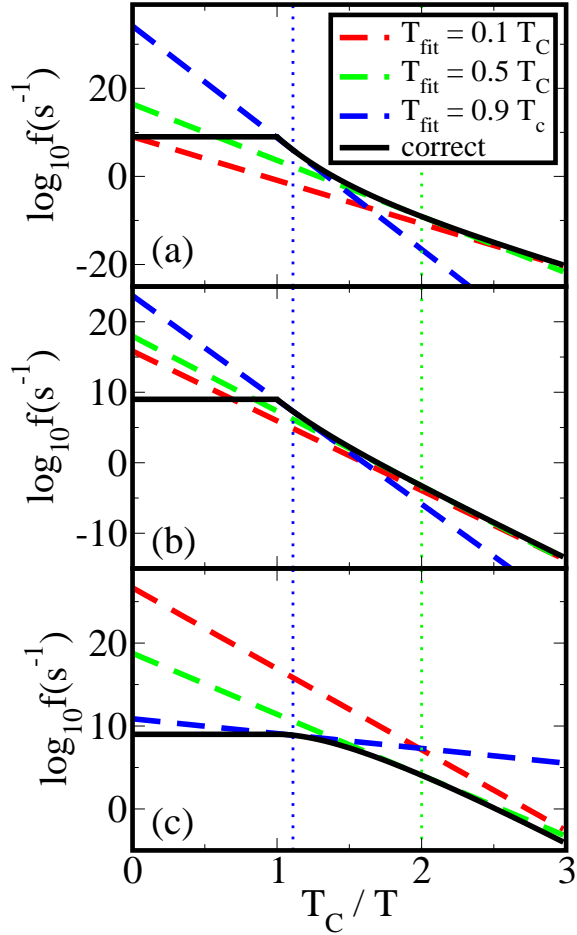}
 \caption{\label{fig::fitvscorrect}Linear fits (dashed lines) to 
the switching rate $\lg f= \lg f_0-\lg (e) K_0M(T)^2/(k_B T )$ (full line),  
performed in a finite temperature interval around the given $T_{\rm fit}$
indicated by the vertical dotted lines (except for $T_{\rm fit}=0.1 T_C$ which
lies out of bounds) for (a) a Brillouin function for $S=1/2$, (b) a
Langevin function, and (c) a linear dependence (cf.\ Fig.\ 
\ref{fig::mag}). The exponents of $f_0^{\rm fit}= f{\rm fit}(T_C/T \to 0)$ can differ 
from the actual $f_0$ by several
orders of magnitude. 
The parameters are $f_0=10^9$ s$^{-1}$, $K_0/M(0)^2=1$ eV and $T_C=500$ K.}
\end{figure}

For the three cases of $M(T)$ used above, the deviation of the prefactor
becomes also temperature dependent now (this is not the case for the
simplified model used in the above paragraph). This effect can be very
pronounced. To show this, we use the same model
parameters as above and numerically perform linear fits according to Eq.\ 
(\ref{eq::ffit}) for a given small temperature interval $T\in [T^{\rm
fit}-T_C/200,T^{\rm fit}]$ in order to mimic the situation of typical
experiments. 
\begin{figure}[t!]
 \centering
     \includegraphics[width=.55\textwidth]{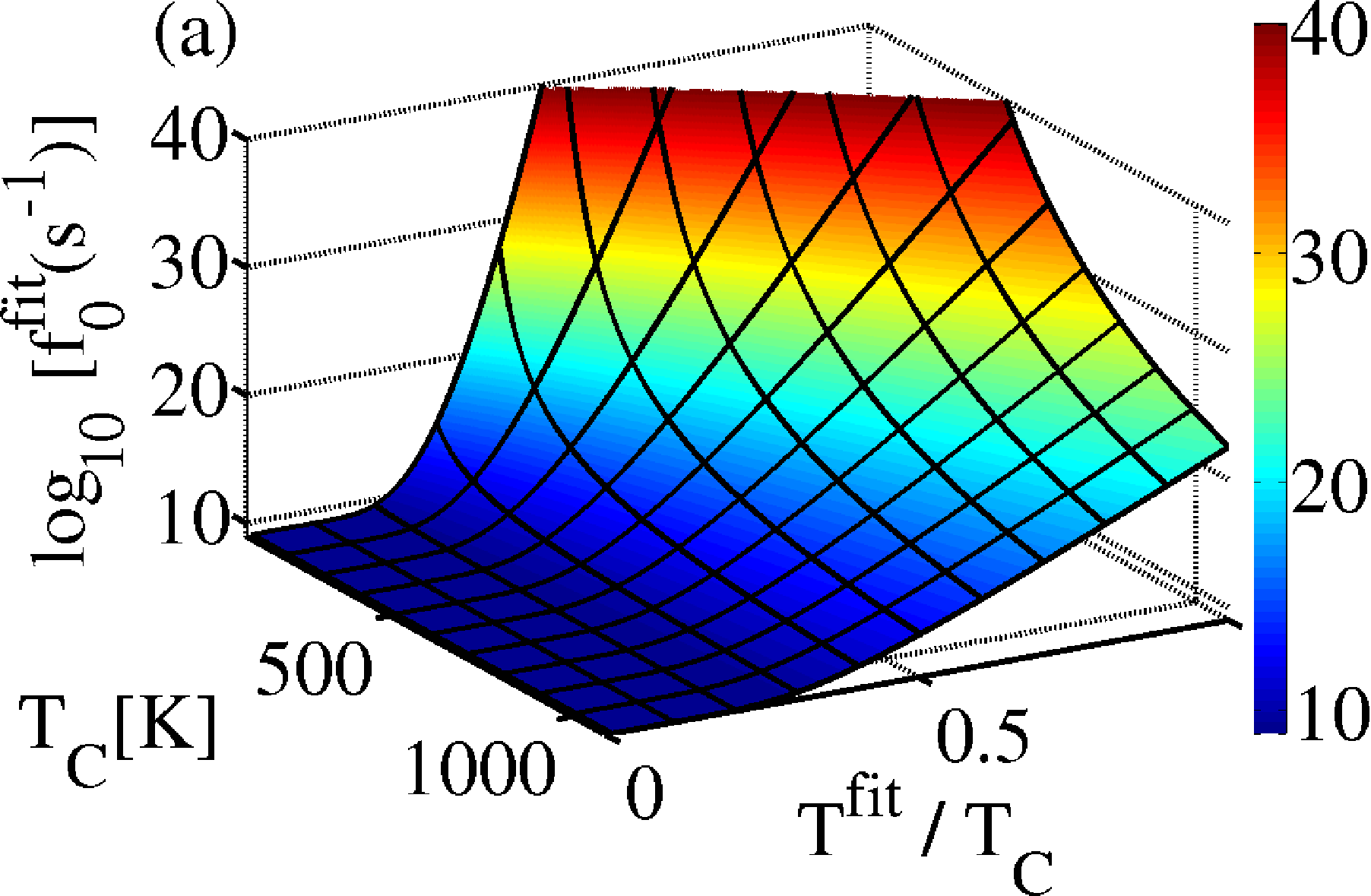}
     \includegraphics[width=.55\textwidth]{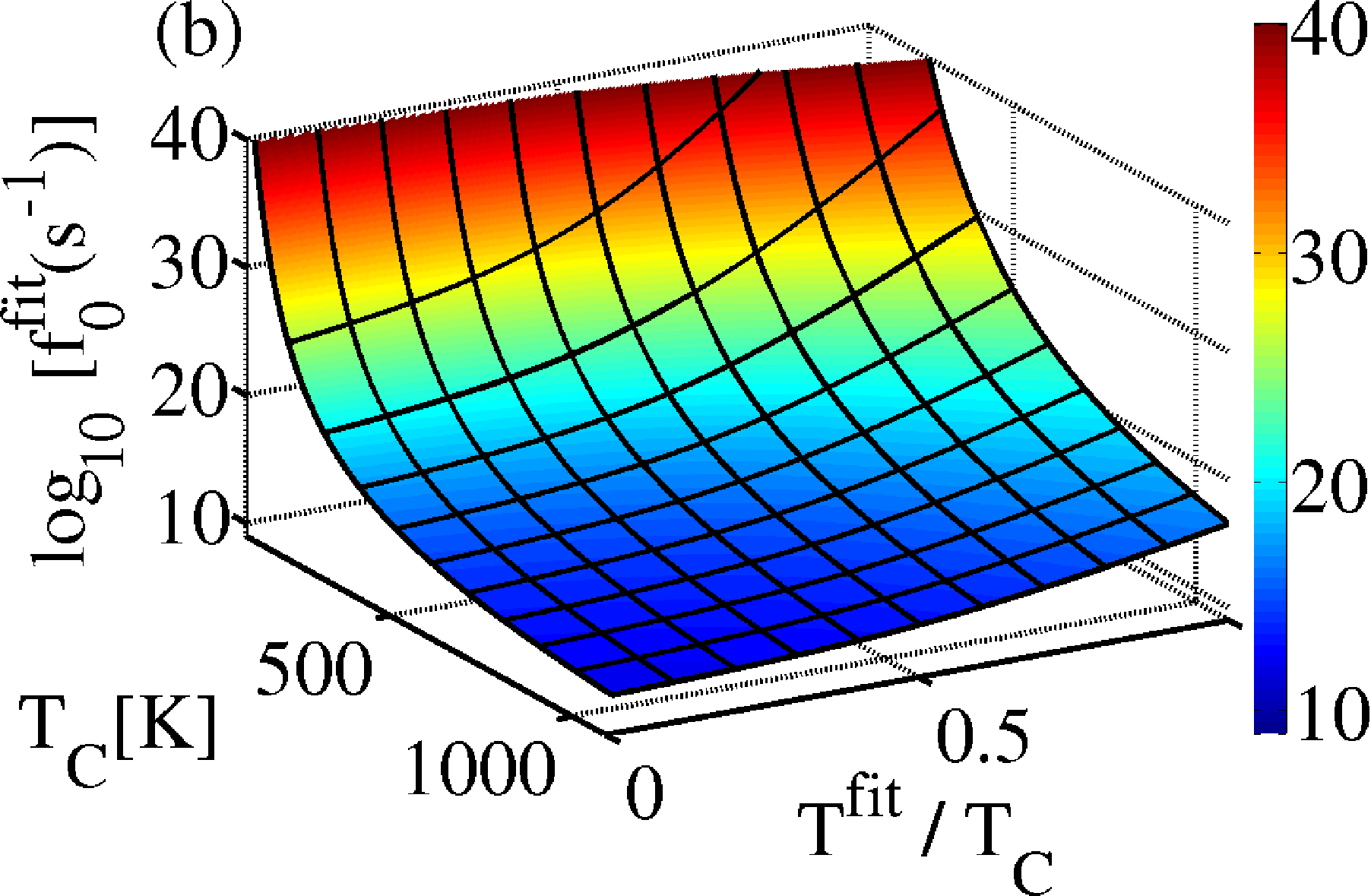}
     \includegraphics[width=.55\textwidth]{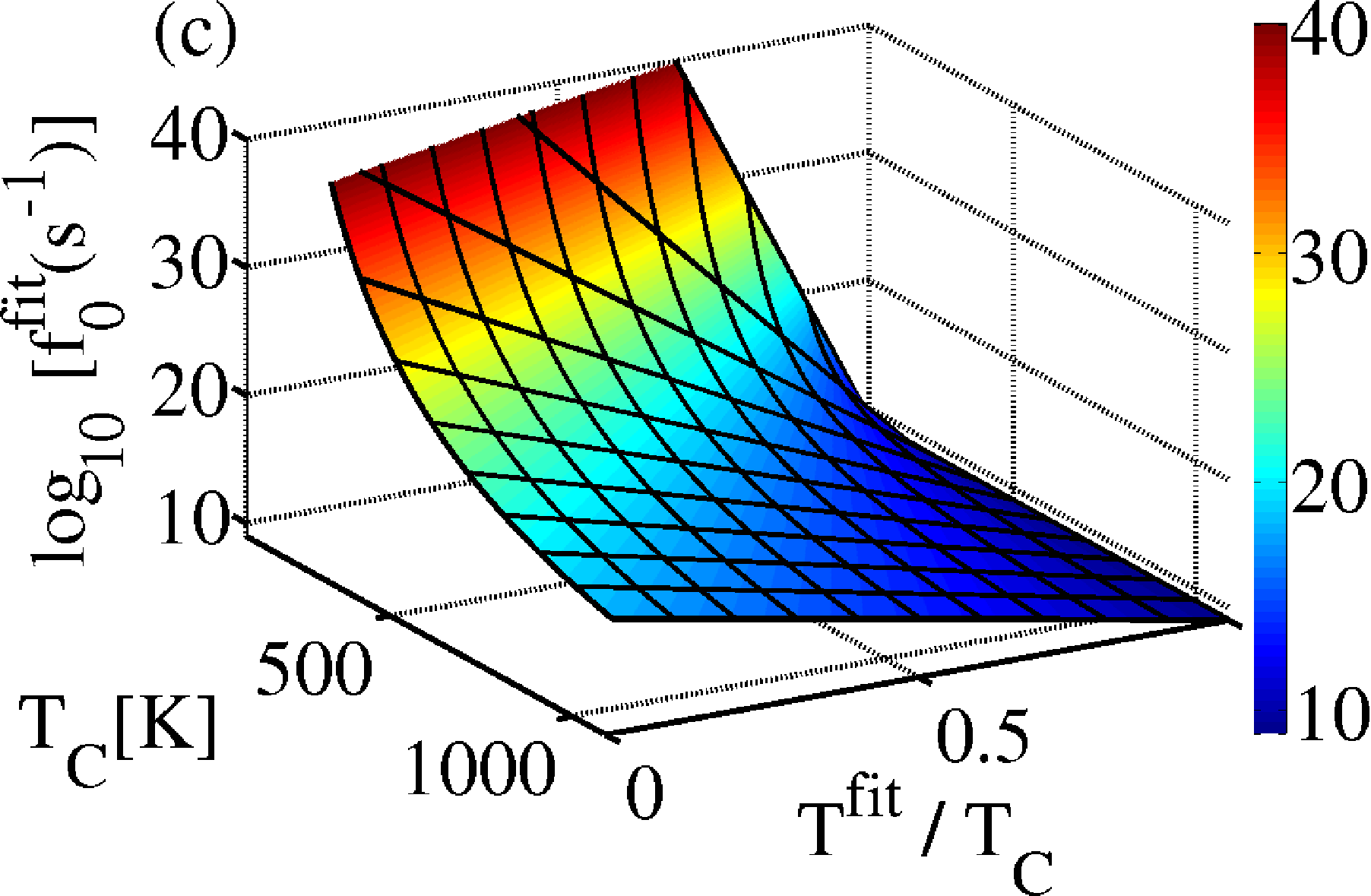}
     \caption{\label{fig::f0exp}Fitted exponent of the attempt
frequency. The magnetization is calculated by (a) a Brillouin function for
$S= 1/2$, (b)
the Langevin function (classical magnetic moment) and (c) by a linear
approach (cf. Fig. \ref{fig::mag}). The correct exponent would be $\lg
f_0({\rm{s}}^{-1})
= 9$.} 
\end{figure}
The results are shown in Fig.\ \ref{fig::fitvscorrect}. The linear fits in
general overestimate $f_0$ by several orders of magnitude. Interpreted as
attempt frequencies, such linear fits yield unrealistically large values.
This explains the unphysical values of the attempt frequencies reported in the
analysis of experimental data on the basis of an Arrhenius or a Vogel-Fulcher
law \cite{krause2009magnetization,tadic2014magnetic,neumann_phd,
tackett2010evidence}.Naturally, the fitted values of $f_0$ are more accurate when the magnetization changes only very little in the respective temperature interval used for the fit. Then, the magnetization is well approximated by the
constant $K_0^{\rm{fit}}$. For a Brillouin function, the (squared) magnetization varies
less at lower temperatures (cf.\ Fig.\ \ref{fig::mag}). Thus, the
best fits are achieved for very low temperature. On the
contrary, $m(T)^2 = [1-(T/T_C)]^2$ changes the least close to the Curie temperature,
which is also reflected in Fig.\ \ref{fig::fitvscorrect}. Obviously,
fits with the assumption of a constant energy barrier cannot deliver  
reliable results for $K_0$, when $\Delta E(T)$ is actually changing with $T$.
Even though $K_0$ probably is not misdetermined by orders of magnitude as it
is fitted directly (and not its logarithm at
$T\to\infty$ as for $f_0$), the correction of the barrier height has to be
taken into account as well. 

The important consequences of a finite temperature window chosen for the
analysis for the exponent of the fitted prefactor are illustrated in more detail
in Fig.\ \ref{fig::f0exp}. We vary the borders of the temperature window used
for the fits and also the Curie temperature. In general, the regions for the
optimal temperature $T^{\rm{fit}}$ are confirmed. An almost
temperature-independent magnetization 
yields a well fitted attempt frequency $f_0^{\rm{fit}}\approx
f_0=10^{9}$ s$^{-1}$. Everywhere else away from this optimal value, it is
significantly overestimated. Moreover, the lower the Curie
temperature is, the worse are the fitted results, which can be again
traced back to the temperature dependence of $M(T)$. As the derivative
$\partial_{\tilde T} M^2(\tilde T)$ with respect to $\tilde T=T/T_C$ remains
constant for all $T_C$ for the three cases treated here, the 
derivative $\partial_T M^2(\tilde T)=T_C^{-1}\partial_{\tilde
T} M^2(\tilde T)$ is proportional to the inverse Curie temperature. Thus, we find
 a larger deviation for a lower $T_C$. This is important for the experimental
analysis since superparamagnetic phenomena are typically investigated for 
very small samples where the Curie temperature may be notably smaller than in a
bulk sample \cite{schneider1990curie}. 


\section{Conclusions}
In this work we have illustrated quantitatively that a temperature-dependent magnetic anisotropy in 
a superparamagnetic system can give rise to significant corrections to the
standard N{\'e}el-Brown-Arrhenius law with a temperature independent energy
barrier. For a correct interpretation of the parameters, the energy barrier for
the switching behavior of the collective magnetic moment has to be based on a
temperature-dependent magnetic anisotropy even in a simple model. This generates
corrections to the N{\'e}el-Brown-Arrhenius law. In particular, the prefactor can
no longer be determined in general by a fit of measured data on the basis of a
temperature-independent anisotropy. Even if the fit appears plausible
in a given finite temperature range where a seemingly linear dependence of $f$ on
$1/T$ is found, the attempt frequency can be incorrectly determined by several orders of
magnitude. 

Although the present discussion applies to the mentioned superparamagnetic
systems, the conclusions are general and fundamental. In many other physical
situations, an Arrhenius-like switching behavior is found on the basis of
assuming a temperature-independent energy barrier. In particular when this
energy barrier is the result of a collective statistical behavior of many
constituents, it may contain an intrinsic temperature dependence which has to
be carefully taken into account, in particular when the prefactor is
interpreted in terms of an attempt frequency. 
\section*{Acknowledgements}
We acknowledge support from the DFG Sonderforschungsbereich 668 ``Magnetismus
vom Einzelatom zur Nanostruktur''.

\end{document}